\documentclass[usenatbib]{aa}
\usepackage{graphics,color,latexsym} 
\usepackage{times}

\usepackage{natbib}
\usepackage{amssymb}
\usepackage{epsfig} 
\usepackage{scrtime}
\usepackage{multirow}
\usepackage{txfonts}
\usepackage{lscape}

\def\mean#1{\left< #1 \right>}

\begin{document}

\title{Dependence on the environment of the abundance function of light-cone simulation dark matter haloes}
\author{Maria Chira \inst{1,2} \and Manolis Plionis \inst{1,2} \and
  Pier-Stefano Corasaniti \inst{3} }
\institute{Physics Dept., Aristotle Univ. of Thessaloniki, Thessaloniki
  54124, Greece \and
National Observatory of Athens, Lofos Nymfon, 11852 Athens, Greece \and
LUTH, UMR 8102 CNRS,Observatoire de Paris, PSL Research University,
Universit\'e Paris Diderot, 92190 Meudon, France}

\date{Accepted 16/5/2018}

\abstract
{}
{We study the dependence of the halo abundance function (AF) on
  different environments in a whole-sky $\Lambda$CDM light-cone halo catalogue
extending to $z\sim 0.65$, using a simple and well-defined halo
isolation criterion.}
{The isolation status of each individual dark matter halo is
  determined by the distance to its nearest neighbour, which defines
  the maximum spherical region devoid of halos above a threshold mass around it (although the true size
  of such region may be much larger since it is not necessarily spherical). A versatile double power-law Schechter
  function is used to fit the dark matter halo AF, and its derived parameters are studied as a function of halo
  isolation status.}
{ (a) Our function fits the halo abundances for all halo
  isolation statuses extremely well, while the well-established theoretical mass
  functions, integrated over the volume of the light-cone, provide an
  adequate but poorer fit than our phenomenological model.
 (b) As expected, and in agreement with other studies based on snap-shot
simulations, we find significant differences of the halo abundance
function as a function of halo isolation, indicating different rates
of halo formation. The slope of the power law and the characteristic mass of
the Schechter-like fitting function decrease with isolation, a result
consistent with the formation of less massive haloes in lower density
regions. (c) We find an unexpected upturn of the characteristic mass of the most
isolated haloes of our sample. This upturn originates and
characterises only the higher redshift regime $(z\gtrsim 0.45),$ which probably
implies a significant and recent evolution of the isolation status of
the most isolated and most massive haloes.}
{}
\keywords{cosmology: dark matter -- N-body simulations}

\authorrunning{M. Chira et al.}
\titlerunning{Dependence of DM haloes on the environment}

\maketitle

\section{Introduction}
According to the cold dark matter (DM) paradigm, cosmic structures
  form hierarchically as a result of the growth of primordial
  density perturbations. The resulting fundamental non-linear cosmic
  structures are known as {\em \textup{dark matter haloes}}, in the
  interior of which baryonic matter collapses to form galaxies, groups
  of galaxies and clusters of galaxies.

It is well established that the distribution of cosmic structures is far from
uniform. DM haloes,  and consequently, the visible objects they
  host, form a hierarchy of cosmic structures from pairs to
  superclusters of galaxies, constituting what has been called the
``cosmic web'' by \citet{Bond1996}, which has revealed a wealth of different
environments. A fundamental property that has emerged
from observations as well as from N-body simulations, and which is
environmental in its essence, is that they are more clustered
in comparison to the underlying mass
fluctuations. This property is called bias and is explained as being the
result of structures forming at the peaks of the initial random
Gaussian density field \citep[e.g.][]{Kaiser1984,Peacock1985}.

The importance of the cosmic environment was first provided through 
indications for an environmental dependence of galaxy
properties by \citet{Dressler1980}, who showed that
galaxy Hubble type and ambient galaxy density are tightly
correlated. Since the unveiling of the cosmic web, interest has been
growing, and systematic studies have been conducted to quantiyfy
  and explain the environmental dependencies of galaxy
  properties.

Recent studies have shown that the environment correlates not only with
various galactic properties
\citep[e.g.][]{Gomez2003,Boselli2006,Blanton2007,Croton2007,
Forero-Romero2011,Eardley2015,Metuki2015} , but also with the
 properties of the DM halo within which
they reside
\citep[e.g.][]{Navarro1997,Bullock2001,Schuecker2001,Plionis2002,
Wechsler2002,Sheth2004,Gao2004,AvilaR2005,Gao2005,Zhu2006,Harker2006,
Wechsler2006,Gao2007,Martinez2006,Ragone-Figueroa2007,Libeskind2011,
Libeskind2012,Libeskind2013,Libeskind2014,Lee2017}. 
This indicates that quantifying
the inter-relation between halo properties (e.g. shapes,
accretion rates, spin parameters, alignments,
substructure, and formation times) and the environment,
local and large scale, can shed light onto the structure formation processes.

A crucial point that has emerged in all studies is the definition of
the \textup{environment} itself. A large variety of methods has been
used to quantify the effect of the environment on the distribution of
  galaxies and DM haloes 
\citep[for an overview, see][]{Muldrew2012,Libeskind2018}. As an example, some of the
  works have used a nearest-neighbour approach,
  while others have defined the ambient density field after applying a
  variety of smoothing kernels to the point-like distribution of
  haloes.

One of the many properties of the galaxy and halo distributions that seems
to correlate with the environment is the halo mass function, the study
of which is crucial in order to develop an understanding of galaxy and
structure formation processes. Pioneering work on these issues is that
of \citet{Press1974}, extended by \citet{Bond1991} to
  include the excursion set formalism and by \citet{Sheth2001} to
  include the more realistic ellipsoidal collapse model
  which takes into account the triaxiality of the Gaussian density
  field perturbations \citep{Doroshkevich1970,Bardeen1986}. The
  resulting DM halo mass
  functions were further improved by many other studies
\citep[e.g.][]{Sheth1999,Jenkins2001,Warren2006,Reed2007,Valageas2009,
Tinker2010,Ma2011,Corasaniti2011}.

Furthermore, other possible
environmental dependencies of the halo and galaxy properties 
could be an ingredient of structure formation processes and
thus many of the studies cited previously have investigated such dependencies. 
A quite common result of such studies is 
the higher abundance of massive haloes in dense environments \citep[e.g.][]{Lemson1999,Maulbetsch2007}.

Thus the abundance of haloes of different mass seems to differ in different
environments. However, it is not yet clear if this effect depends on the
ambient density or on the web-element type (knots, 
filaments, sheets, or voids). \citet{Hahn2007} reported a variation in halo mass function with web-element
classification, in the direction of an increasing higher mass end of the
halo mass function at the upper part of the web-element sequence,
 which corresponds to areas with higher over-densities \citep{Hoffman2012,Libeskind2015}. 

Two interesting and related studies  recently
reached apparently contradicting results. \citet{Alonso2015} reported 
that the halo mass function does not depend on the
web-element type, but only on the local density, while 
\citet{Metuki2016} found the opposite. As suggested in the latter work, a possible
explanation for this discrepancy is the fact that \citet{Alonso2015} 
defined the local density through a constant radius-smoothing kernel, 
while \citet{Metuki2016} used an adaptive kernel that
explicitly depends on the virial radius of the haloes. The contradicting 
results of \citet{Alonso2015} and
\citet{Metuki2016}  reveal that the relation
between environment and halo properties is still open to
discussion and demands further investigation.

We have chosen a different approach
to define the environment here that is based on a simple and clear-cut halo isolation
criterion. We focus on haloes of a mass that would today host
groups and 
clusters of galaxies. We chose to examine the behaviour of the halo
abundance for isolated haloes within a specific radius, and similarly,
of pairs of haloes, and compare it with the behaviour of less isolated haloes. The
definition of environment in terms of isolation rather than in terms
of specific values of the local density field  or in terms of a
  web-element classification is also interesting 
from the observational point of view and can relatively 
easily be applied to redshift surveys.

\section{Simulation data}

We used halo catalogues of light-cone data generated on flight during
the realization of a subset of N-body simulations from the "Dark
Energy Universe Simulation" (DEUS) project
\citep{Alimi2010,Rasera2010,Courtin2011} 
that are publicly available through
the DEUS database\footnote{www.deus-consortium.org/deus-data/}. The
N-body runs have been performed using the adaptive mesh refinement
code RAMSES, which is based on a multigrid Poisson solver
\citep{Teyssier2002,Guillet2011} 
for Gaussian initial conditions
generated using the Zel’dovich approximation with the MPGRAFIC code
\citep{Prunet2008} and input linear power spectrum from CAMB
\citep{Lewis2000}. 
The light-cone data used here are from
simulations of 2592 Mpc/h boxlength with $2048^3$ particles for a standard
$\Lambda$CDM model with parameters calibrated against supernova Type Ia
from the UNION dataset \citep{Kowalski2008} and measurements of the
cosmic microwave background anisotropies from the Wilkinson Microwave
Anisotropy Probe (WMAP) 5-year data \citep{Komatsu2009}, that
is,
$\Omega_m=0.267$ and $H_0=100 h$ km s$^{-1}$ Mpc$^{-1}$.

The light-cone halo catalogue covers the full sky out to a redshift
$z<0.65$. The haloes contain more than 100 particles, while the
particle mass resolution is $m_p=1.5\times
10^{11}\,M_{\odot}/h$. Haloes were detected in the light-cone
using the code pFoF, a Friend-of-Friend halo finder \citep{Roy2014}. 
The total number of haloes in our catalogue is $\sim
3.15\times 10^6$.

As a final note, we wish to stress that
  light-cone simulation data are extremely useful for testing 
  algorithms and methodologies in order to enable direct
  comparisons with observational redshift data. Our particular
  simulation has the halo mass limit and volume traced, which makes it
  suitable for large statistical studies of the
  abundance and physical properties of massive haloes hosting 
groups and clusters of galaxies.
  This is indeed useful, since such objects will be studied
  in future cluster surveys, provided by {\sc eRosita} 
\citep[e.g.][]{Hofmann2017} in the X-ray and {\sc LSST} 
\citep[e.g.][]{Marshall2017} and {\sc Euclid}
\citep[e.g.,][]{Sartoris2016} 
in the optical.

\section{Method}
\subsection{Definition of local environment}
We here use a rather simple approach to define the local
environment of dark matter haloes that is especially tailored to reveal the
inter-halo dynamics. We avoid the approach of categorising the
different regions of the cosmic web according to a range of
web-elements (knots, filaments, sheets, and voids), and we use a criterion
that is centred on each individual halo.
In detail, we identified the nearest neighbour of
each halo and its corresponding distance, which we call 'isolation'
radius, $R_{\rm isol}$. Our criterion resembles that of
\citet{Haas2012}, 
although it is very different in that we
defined the distance regardless of the halo mass ($M>10^{13} M_{\odot}$). 

 The isolation radius defines a spherical region that is devoid of
 other haloes. Therefore, in the usual jargon, a very small
isolation radius corresponds to a high-density region, while a large
isolation radius corresponds to an underdense region \citep[see the review by][]{vdWeygaert2016}.  
We note, however, that the true volume of such a 'void'
may be much larger as it may extend considerably towards directions other than that of the nearest neighbour.
We also note that although our halo catalogue contains haloes with $M\ge
10^{13} M_{\odot}$, we used as central haloes (around which we
defined the local environment) only those with 
$M\ge 2.5 \times 10^{13} M_{\odot}$. This is required in order to be
able to define 'isolation' towards lower mass haloes as well
and thus
ensure that our results
are not heavily biased by the mass limit of our simulation halo data.  
\begin{figure}
\centering
\includegraphics[scale=0.7]{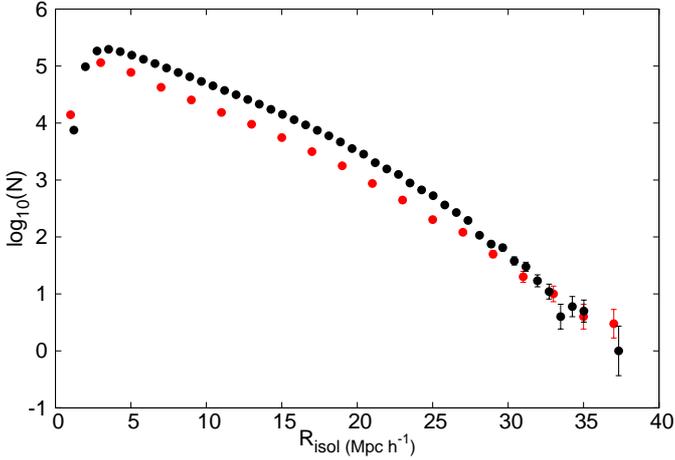}
\caption{Frequency distribution of the isolation radius: for the
    complete sample of light-cone DM haloes with $M\ge 2.5 \times
    10^{13} M_{\odot}$ (black points) and for the sample of close pairs of DM haloes
    with separations $\leq 2.5 h^{-1}$ Mpc (red points). Error
bars correspond to Poisson uncertainties.}
\label{risol}
\end{figure}

In Figure \ref{risol}, black points correspond to the frequency distribution
  of $R_{\rm isol}$ for all haloes in the light-cone simulation, from
which it is evident that the values of $R_{\rm isol}$ span a wide range
from $\sim$0.85 to 37 $h^{-1}$ Mpc. Evidently, the large majority
  of haloes have close neighbours, as expected from the hierarchical
  clustering scenario. However,
  the existence of extremely isolated haloes residing in huge
  underdense regions is particularly interesting; some of them
are quite massive. We have found massive haloes ($M>10^{14} M_{\odot}$) with isolation
  radii of up to 20-30 $h^{-1}$ Mpc. This can
  give insight into halo formation processes in extreme environments.

Although haloes with small $R_{\rm isol}$ should be
considered as residing in high-density regions, we cannot
exclude the possibility of isolated pairs of haloes. Selecting
the central DM haloes with $R_{\rm isol}< 2.5 h^{-1}$ Mpc and using
the distance to the second nearest neighbour as a further isolation
criterion, we find that there are many pairs of DM haloes (central halo
plus its first neighbour)  with different isolation status, some are
even found within regions of radii as large as $\lesssim 38 \; h^{-1}$
Mpc that are devoid of other haloes within the mass limit of our catalogue.
In the same figure we show with red points the frequency
  distribution of the isolation radius
for these close pairs of DM haloes, which also shows the wide
range of isolation statuses of halo pairs.

A final but important methodological issue is related to the fact that
by directly using the $R_{\rm isol}$ parameter as a characterisation of
local environment, we do not take the halo size into account, which
unavoidably affects the available minimum separation among different
halo sizes. Therefore, we chose in the remaining to use a
parameterised
characterisation of the local environment provided by the isolation
radius in units of the halo virial radius,
$R_{\rm isol}/r_{\rm vir}$. This definition is in accordance with the
definition of local density introduced in
\citet{Metuki2016},  which is more reliable than the
approach of \citet{Alonso2015}, since the
use of an adaptive radius smoothing kernel avoids the problem of
underestimating the density around less massive haloes when the density
is calculated at a fixed radius. Thus,
the method of adaptive aperture is a way of normalising
density with the virilization properties of each halo.

\subsection{Halo abundance function}

 As described above, the main goal of this work is to study the environmental 
dependence of DM halo abundances using a simple but
novel environmental isolation criterion.  We note that we
use the term "abundance function", $N_{AF}(M)$, instead of "mass
function" in order
to highlight that we are applying our analysis on light-cone data,
meaning that our catalogue has a wide span in redshift, in contrast to
the "traditional" mass function, which is defined for specific redshifts. 
In order to 
quantify $N_{AF}(M)$ for the different isolation status, it is
essential to identify a versatile and relatively simple analytical
function to fit the simulation halo abundances. 
Since the theoretically motivated  $\Phi(M,z)$'s, usually based on the
Press \& Schechter formalism, reflect the whole population of haloes at a given redshift independent of
 their location, there is no direct way of applying them to the
  halo distribution in different environments  unless one allows the
  numerical parameters to be fitted directly by the halo
  data in each different environment (see discussion relevant to
  Figure 4). Moreover, our choice of selecting a simple and versatile analytical
  function was also made because we did not wish to
  make a detailed comparison of the great variety of
  theoretical $\Phi(M,z)$ in order to select an "optimum"
  model. Such a comparison has been performed in other studies
  \citep[e.g.][]{Watson2013} and is beyond the scope of the current work.

We thus identified a useful quantification of the DM halo
abundances, based on a Schechter-like function, which is known to
represent the luminosity function of galaxies accurately. The idea
behind this choice is the expectation that a similar functional
form can be expected to represent the DM halo abundances sufficiently
well through the mass-to-light relation. A Schechter-like function, 
with a double power-law, was found to represent the
abundances of our DM haloes accurately. We note that the second power-law,
necessary to fit the high-mass end of the overall AF, is not necessary
when we consider DM haloes of medium and high isolation.

The functional form of the abundance function that we used is
\begin{equation}
N_{AF}(M)= \left[C_1\left(\frac{M}{M_{\star}}\right)^{\alpha} + 
C_2\left(\frac{M}{M_{\star}}\right)^{\beta}\right]
\exp\left(-\frac{M}{M_{\star}}\right)
\label{eqn_Schechter}
,\end{equation}
where $C_1$ and $C_2$ are normalisation factors related to the halo
number density, $M_{\star}$ is the characteristic mass related to the
knee of the abundance function, and $\alpha$ and $\beta$ are the
exponents of the power laws.

The halo abundance as a function of mass was measured in logarithmic
mass bins of width $\delta\log M\simeq 0.0693$, which is a
compromise among the different subsamples that we used, in order that the
halo numbers, in the most under-abundant bins, are not dominated by
Poisson errors. The analytic fit to the resulting halo abundances was
performed using the usual $\chi^2$ minimization procedure:
\begin{equation}
\chi^2({\bf p})=\sum_{i=1}^N \frac{\left(\log N_i(M)-\log N_{AF}(M,{\bf
    p})\right)^2}{\sigma_i^2}
,\end{equation}
where $N_i$ is the number of haloes in the i$^{th}$ mass bin,
$\sigma_i$ is the uncertainty in $\log N_i$ , for the
  calculation of which we chose to consider the uncertainty in $N_i$
  equal to $3\sqrt(N_i)$. We note that we also used bootstrap
  uncertainties with no effect at all on our results.
The sum is over the halo mass bins, and  the vector ${\bf p}=(\alpha,
\beta, M_{*}, C_1, C_2)$ contains the free parameters.
The specific procedure that we used entails the following steps:
\begin{itemize}
\item We first fit the single power-law Schechter-like function and
  determined the best-fit values of $\alpha$, $C_1$ and
  $M_{\star}$. 
\item We then fit the double power-law Schechter-like function, but 
  keeping the above three parameters constant to their best-fit values
  of the first step, that is, allowing only $C_2$ and $\beta$ to be fitted in
  this second step.
\item If the reduced $\chi^2$ provided by the fit of the second step
  was lower than that of the first step, we considered that the double
  power-law version of the Schechter-like function is a better
  approximation to the halo abundance function under study.
\end{itemize}
 We also tested an alternative procedure by forcing the
  $C_2$ parameter to be fitted over a restricted range of low
  values, to take into account the small contribution of the
  corrective term, and then allowing all five free parameters to be
  fitted simultaneously, which led to results that were
very similar to those of the previously described procedure,
however.

\begin{figure}
\centering

\includegraphics[scale=0.7]{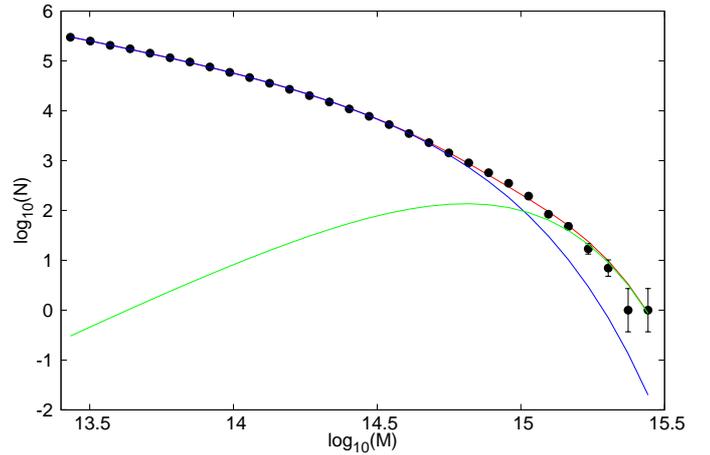}
\caption{Abundance of haloes with $M\ge 2.5\times 10^{13}
  M_{\odot}$. The analytic Schechter-like function fits, 
  $N_{AF}(M)$, are represented
  by continuous curves. The single power-law Schechter function is
  shown in blue, the double power-law function in red, and in
  green we separately show the second power-law fit.}
  \label{fit_af}
\end{figure}

As a manifestation of our procedure, we present in Figure \ref{fit_af}
the overall halo-abundance function of our complete light-cone halo
sample (circular
points) and the best-fit Schechter-like functions. The blue line
corresponds to the single power-law fit (first step), where it is
evident that it represents a wide dynamical range in mass well, except
for the highest-mass regime. The green line
corresponds to the second power-law fit, while the joint two power-law
Schechter function of Eq.(1) is shown as the red curve. The excellent
fit of the latter to the data is evident. We note that with our approach, the
second Schechter-like function is used only as a small correction to the
main fit, the parameters of which remain unchanged, meaning that the value of
$M_{\star}$ that is used in the second Schechter-like function is
fixed to the value determined by the fit to the initial single
power-law Schechter function.

\begin{figure}
\centering

\includegraphics[scale=0.7]{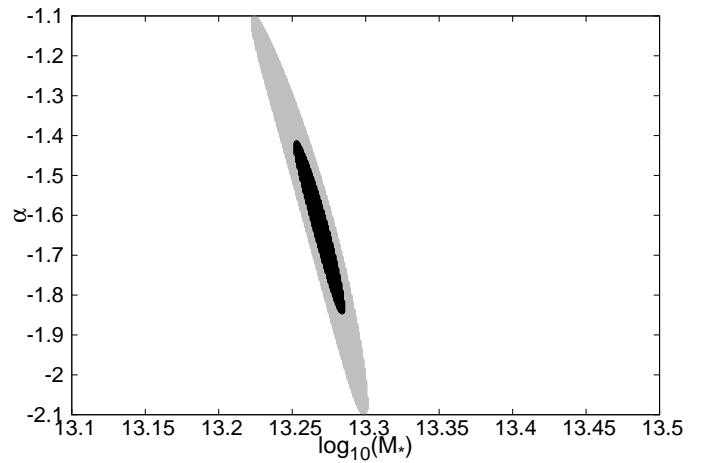}
\caption{1$\sigma$ (black) and 3$\sigma$ (grey) contours in the
  $\alpha$-$M_{\star}$ parameter space.}
  \label{a_and_M}
\end{figure}

In order to investigate possible degeneracies among the parameters, we
plot in Figure \ref{a_and_M} the $1\sigma$ and 3$\sigma$ contours, corresponding
to $\chi^{2}-\chi^{2}_{min}= 2.3$ and $11.83$, in the $\alpha$ and
$M_{\star}$ solution space.
Although there is an
important degeneracy of the $\alpha$ parameter, the
$M_{\star}$ parameter is very well constrained, with an extremely small
uncertainty. This result is typical of all the different halo samples
analysed in this work. The uncertainties of the individual parameters
$\alpha$ and $M_{\star}$ are
calculated after projecting the 1$\sigma$ surface on each parameter
axis and estimating its maximum projected range. This definition
provides a rather (artificially) wide uncertainty
range. Alternatively, one could marginalize one parameter over the
other and then estimate each individual parameter uncertainty, which 
 would then, because of the
significant degeneracy, provide an underestimate of the true
uncertainties, however.

Finally, we tested the reliability of our results also with the
Bayesian-based {\em emcee} \citep{Foreman-Mackey2013} to estimate
the best-fit values of the parameters of our model.  {\em emcee}
is an MIT-licensed Python implementation
of the affine-invariant ensemble sampler for Markov chain Monte Carlo
(MCMC) proposed by \citet{Goodman2010} and it is authored by Dan
Foreman-Mackey. We found that both parameter estimation approaches
give results that agree excellently well.

\begin{figure}
\centering
\includegraphics[scale=0.45]{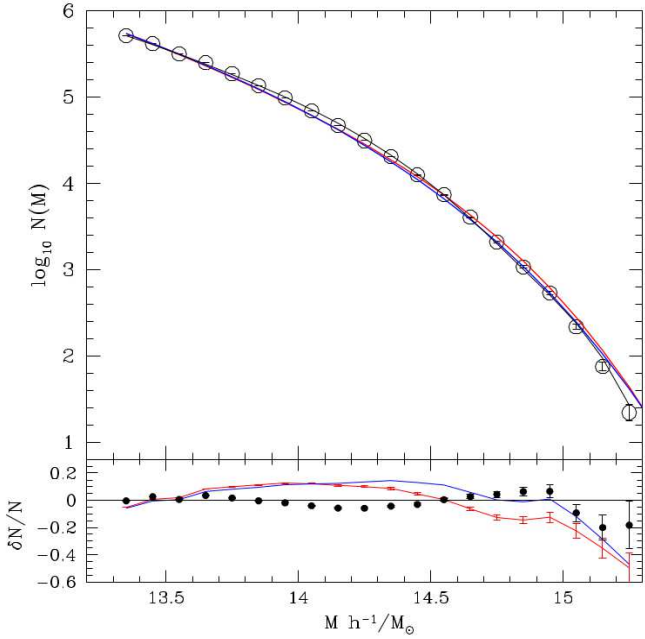}
\caption{Empty points represent the light-cone halo $N(M)$, the
    red curve shows the volume-integrated \citet{Reed2007} $\Phi(M,z)$, the
    blue line shows the \citet{Jenkins2001} $\Phi(M,z), $  and
the black
    curve represents the double power-law Schecther function. The lower panel
    shows the relative deviations of the three models from the data
    $N(M)$, with the filled black points representing the deviation of our
    Schechter-like function. It is evident that the latter functional
    form represents
     the data better. The two theoretical $\Phi(M,z)$ are the best
    fits to the halo data from six models.}
  \label{mf_comp}
\end{figure}

As a further test of our choice, we attempted to fit the abundance
function of our light-cone haloes with various theoretical DM
halo mass function models. To this end, we integrated over redshift 
the theoretically motivated $\Phi(M,z)$, within the volume of the
light cone and found
that the resulting theoretical abundance functions, $N_{th}(M)$,
represent our data quite well. In
Figure \ref{mf_comp} we present the two best-fit models to the halo data
(from six models) and the double power-law Schechter form we
used. The red curve in the figure corresponds
to the volume-integrated \citet{Reed2007} model, the blue line
to the \citet{Jenkins2001} model, and the black curve to the
double power-law Schechter function. It is clear, especially from the
lower panel, where we present the deviations of each model from the
light-cone halo data, that the latter functional form is a much better
fit to the data. This is also supported by the values of the reduced $\chi^2$,
which for the theoretical models are more than an order of magnitude
higher than for our fitting function. 

We wish to add that we have also
allowed the different numerical parameters of the theoretical mass functions  to
be fitted directly by the data, that is, we followed the same sort of
approach as with our Schechter-like function, and although we find
theoretical mass functions for which the obtained values of
$\chi^2$/df are lower that than those
obtained using their nominal parameter values, they are still
significantly higher than the values corresponding to our Schechter fit.

Nevertheless, it is important to clarify that our aim in this
work is not to introduce the Schechter-like form as an alternative to
any theoretically motivated model, but only as
a  reliable quantification of the DM halo abundance function, which
allows us to study its behaviour in different environments.

\section{Results}

\begin{figure}
\centering

\includegraphics[scale=0.7]{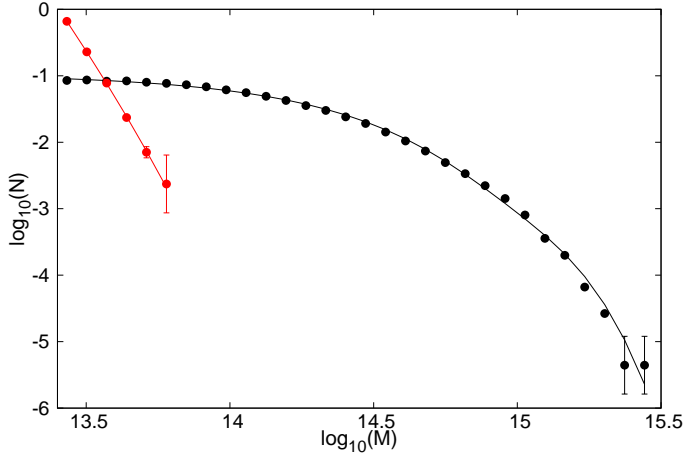}
\caption{ Abundances of haloes of $ R_{\rm isol}/r_{\rm vir} \le 4$
  (black) and $R_{\rm isol}/r_{\rm vir} \ge 50$ and best-fit
  curves. The AFs are normalized to the same total number.}
  \label{af_extreme}
\end{figure}

\begin{figure}
\centering
\includegraphics[scale=0.7]{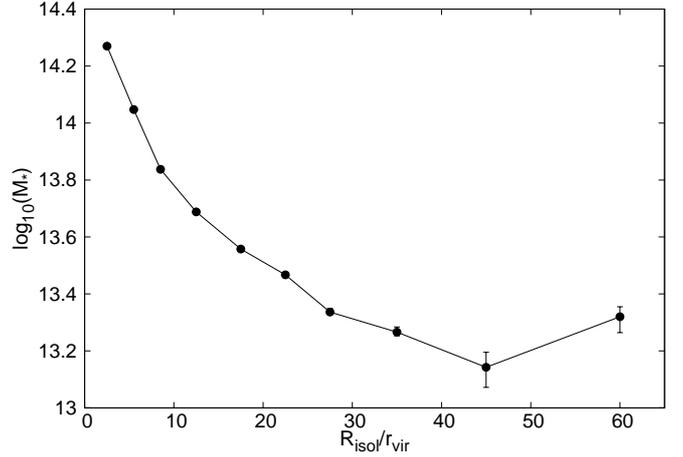}
\caption{Characteristic $M_\star$ parameter, with a
  significantly decreasing tendency except for the last
  bin, which corresponds to the most isolated haloes ($50<R_{\rm
    isol}/r_{\rm vir}< 70$).}
    \label{Mstar}
\end{figure}

In order to realize the main aim of our current work, we
separated our halo catalogue into different subsamples based on their 
$R_{\rm isol}/r_{\rm vir}$ values. This allowed us to
study the differences in the halo abundance function 
for haloes of different
isolation status. The normalized AF, to the same overall number, for
the two extreme cases of isolation status are
shown in Figure \ref{af_extreme}. The two AFs are significantly
different, showing a strong dependence of the AF on the isolation
status, with
the halo AF of highly isolated halo regions being very steep and
dominated by lower halo masses than in dense
regions. For example, the most isolated haloes, corresponding to
underdense regions, have masses that do not exceed $10^{13.75}$
$M_{\odot}$, while haloes in dense regions span the entire
mass interval.
We also estimated the AF in all intermediate values of the halo isolation,
and for each we fitted the function of Eq.(1), extracting the best-fit
parameters $\alpha$ and $M_\star$. We find that
$\alpha$ is a monotonically decreasing
  function of the isolation radius, taking values in the interval $\alpha
  \in [-5.2, -0.2]$, and the behaviour of $M_\star$, as a function
  of the isolation radius is shown in Figure \ref{Mstar}.

The generally decreasing tendency of both $M_\star$ and $\alpha$, 
up to values of $R_{\rm isol}/r_{\rm vir}<50$,
is consistent with what we would expect; the denser the
environment, the more massive the haloes that tend to form. However, we
find an unexpected but statistically significant upturn of $M_\star$
for highly isolated haloes, implying that the most isolated haloes 
tend to be analogously more massive than in less extreme isolation cases. 

In an attempt to understand this unexpected result, we investigated 
the behaviour of the characteristic mass $M_\star$ in two separate
redshift bins to determine whether it shows signs of evolution. We performed our analysis
separately in two redshift subsamples, for which we divided  our total halo
sample into two equal subsamples: one limited to $z<0.456,$ and the other to
$z \in (0.456, 0.625)$.

As a first comparison of the haloes in the two redshift subsamples, 
we discuss the two extreme cases of environments; the densest ($R_{\rm
  isol}/r_{\rm vir}<4$ ) and the most
isolated ($50<R_{\rm isol}/r_{\rm vir}<70$).
We find that the lower redshift subsample contains only $\sim 10\%$ of
 the total DM haloes with $R_{\rm isol}/r_{\rm vir} \in (50, 70)$. We
 verified that this is not due to different comoving volumes probed by
 the two redshift intervals: the corresponding volumes are quite
 similar, $\sim$21.2 Gpc$^3$ and $\sim$27 Gpc$^3$ for the lower
 and higher redshift interval, respectively.
A partial explanation of this difference would be if the DM haloes of the
lower redshift subsample, as a result of further gravitational
evolution, were to tend to have smaller virial radii. 
The mean virial radius in the two subsamples is indeed found to 
be $\mean{r_{\rm vir,1}} > \mean{r_{\rm vir,2}}$, 
while $\mean{R_{\rm isol,1}} \simeq \mean{R_{\rm isol,2}}$, resulting
in lower values of $R_{\rm isol}/r_{\rm vir}$ in the low-redshift subsample. 
A further explanation, which appears to be supported by the
results presented below, is that the largest isolation status
of, especially, massive haloes evolves to smaller isolations at lower redshifts.

\begin{figure}
\centering

\includegraphics[scale=0.7]{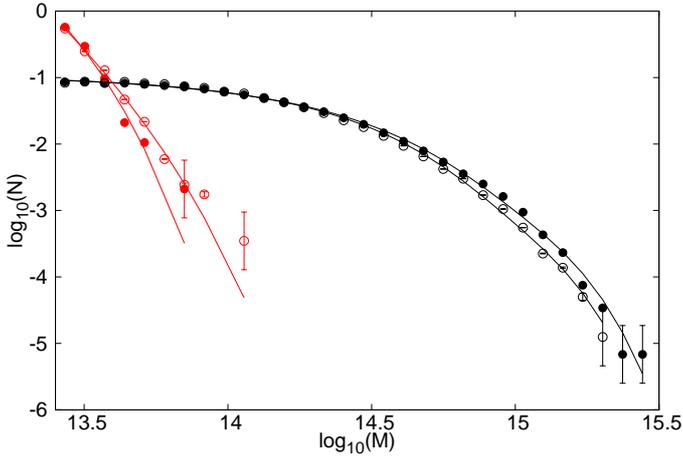}
\caption{Halo abundances for the two extreme different isolation
  status, $1<R_{\rm isol}/r_{\rm vir}<4$ (black) and $40<R_{\rm
    isol}/r_{\rm vir}<50$
  (red). The lower redshift range $z<0.456$ is indicated with filled
  symbols, and the higher redshift range, $z \in (0.456, 0.625),$ with
  open symbols. The AFs are normalized to the same total number.}
  \label{af_extreme_reds}
\end{figure}

In Figure \ref{af_extreme_reds} we compare the halo abundances for
these two extreme cases of environment in the two different redshift intervals. 
Even though we have a relatively small number of highly
isolated haloes and thus statistically important uncertainties, the
normalized AF of the two different redshift intervals shows significant
differences, some of which are expected. Specifically, we find that
for the lowest isolation environment 
(high-density regions), the lower-redshift AF is systematically higher
at the high-mass end than the higher-redshift AF, as expected from the
gravitational
growth of halo mass. However, at the highest isolations (lowest
density regions), the
opposite is true, which might be interpreted 
as meaning that the highest mass and most isolated haloes are more
massive at higher $z$. This interpretation is rather unphysical and
counter-intuitive, while
an alternative, corroborated by the results presented below, is more
probable.

\begin{figure}
\centering

\includegraphics[scale=0.7]{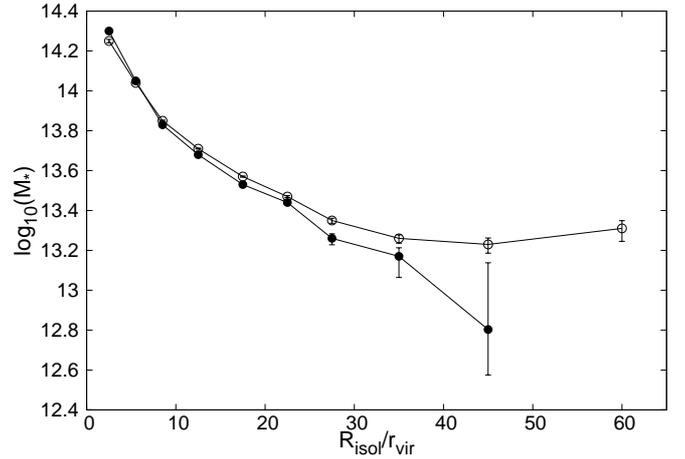} 

\caption{Fitted $M_\star$ parameter of the Schechter-like AF as a
  function of $R_{\rm isol}/r_{\rm vir}$ for the two different
  redshift bins ($z<0.456)$ with filled symbols and $z \in
(0.456,0.625)$ empty symbols). We observe an upturn only for
$R_{\rm isol}/r_{\rm vir}\gtrsim 40$ only in the higher redshift bin.}
\label {Mstar_reds}
\end{figure}

 The redshift-dependent differences were also quantified for all
 different isolation statuses by fitting the
parameters of the abundance Schechter-like function as a function of
$R_{\rm isol}/r_{\rm vir}$. The results of $M_{\star}$ for the two
redshift intervals are presented in Figure \ref{Mstar_reds}.
$M_\star$ shows a consistent decrease with isolation
status in both redshift intervals, except for the most isolated
states, where the upturn in $M_\star$ is present only in the higher
redshift bin. We note that the value of the $M_{\star}$ parameter
  denotes the position of the knee, that is, the mass above which the
  decreasing exponential term dominates. It is not necessary
  to have data over all mass scales to obtain the best-fit value of
  $M_{\star}$, not even scales corresponding to $M_{\star}$ itself. As
  we also explain below, in order to verify that the Schechter
  $N_{AF}(M)$ fits the data well on such occasions, we tested
  some simple alternative forms to find that the Schechter function
  indeed fits the data better, even in cases when the best-fit value
  of $M_\star$ is lower than the mass limit of our catalogue. 

Therefore, we verify that the statistical significant increase of
$M_\star$ for the most isolated DM haloes, with respect to less
isolated ones, is
related to the higher redshift regime and that it disappears at lower
redshifts. This finding, in full agreement
with the results presented in Figure \ref{af_extreme_reds} and discussed previously,
implies that the large isolation of the most massive high-$z$ 
haloes, in other words, the large distance to their
nearest neighbour, decreases with redshift. 
The specific mechanism that causes this behaviour is not
clear and needs further study.

In order to scrutinize these results, especially in the light
  that the most isolated haloes span a relatively narrow range in halo
  mass and thus the range over which we can fit the analytical AF
  function is quite limited, which makes the resulting
  parameters of the Schechter-like function doubtful, we fit
  alternative functional AF forms to the data. In detail, we used
  a power law with two free parameters (the normalisation
  parameter, $C$, and the slope, $\alpha$) and an exponential
  with two free parameters (the normalisation parameter,
  $C$, and characteristic mass, $M_{\star}$). Comparing the resulting
  reduced minimum $\chi^{2}$ values, we found that our original
  Schechter form is the most suitable function to
  also represent the highly isolated halo AF.

\begin{figure}
\centering

\includegraphics[scale=0.7]{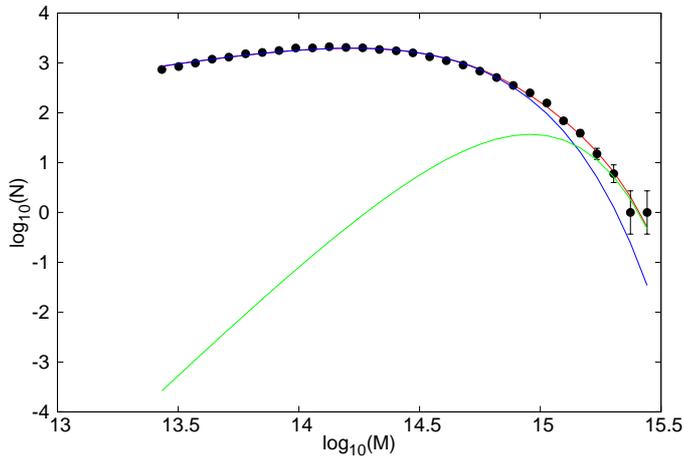}
\caption{AF of haloes that have a first and second nearest-neighbour  with
  a distance  $< 4 r_{\rm vir}$. The analytic Schechter-like function
  fits, $N_{AF}(M)$, are represented
  by continuous curves. The single power-law Schechter function is
  shown in blue, the double power-law function in red, and in
  green we separately show the second power-law fit.}.
  \label{knots_compare}
\end{figure}

Finally, we wish to compare our results with those of other studies
that used more complex and multiparameter procedures to determine the
environment (with all the advantages
and disadvantages they may have): e.g., those of \citet{Metuki2016}
and \citet{Alonso2015}. 
Such a comparison should be made for the range of environments that 
can be identified as equivalent. These are the ``knots'',
corresponding to the highest density
semi-virialized cluster regions \citep[e.g.][]{Metuki2016}. In our case,
we constructed the AF for DM haloes with isolation radii of the first
and second nearest neighbours (see section 3.1) $< 4 r_{\rm vir}$.
The AF we derived is shown in Figure \ref{knots_compare} and indeed has the characteristic
downturn for masses $\lesssim 10^{14} M_{\odot}$, which is in
qualitative agreement with the downturn described by
\citet{Metuki2016} and \citet{Alonso2015} and also with what has been found from
  some of the web-element finders (T-web, V-web, and CLASSIC), 
which are compared in \citet{Libeskind2018} for knots (see their Figure 6,
  top left panel).
We note that the specific lower mass limit of our DM halo sample ($2.5 \times 10^{13}
M_{\odot}$) does not allow us to probe the observed upturn of the mass
function towards lower halo masses described in
\citet{Metuki2016} and \citet{Libeskind2018}, 
a mass range where the latter works
  differ from \citet{Alonso2015}. 
Moreover, a detailed comparison of our results 
with these works is not possible because the size of
the simulation and its resolution are significantly different.

Using the extended isolation criterion (that also
includes the isolation of the second nearest neighbour and imposes that it lies within the
same range as that of the first nearest neighbour) and repeating our
main analysis of examining the behaviour of fitted
parameters $\alpha$ and $M_{\star}$ as a function of isolation status, 
we found no significant change compared with our original
analysis. The two free parameters have exactly the same tendencies,
except for a systematic shift of the $\alpha$ parameter towards
higher values, by $0.498$ on average.
    
\section{Conclusion}
We studied the dependence of the halo abundances on
environment of a light-cone $\Lambda$CDM halo catalogue extending to
$z=0.65$, based on the DEUS simulation project. 
We used a distinct environmental criterion 
centred on each halo with $M>2.5 \times
10^{13} M_{\odot}$ by defining an
isolation region around it within which no other $M>10^{13} M_{\odot}$
halo can be found. A similar isolation criterion can easily be
applied
to observational data, which enables a direct comparison, with minimum
assumptions, between simulations and observations.
Our basic results are summarized below.
\begin{itemize}
\item The halo mass abundances depend strongly on the isolation
  radius, a result similar to that of many other studies that have defined the
  environment with a variety of multiparametric methods.
\item A double power-law Schechter-like function fits  the halo
  abundance of light-cone DM haloes for all isolation radii very well,
  although the second
  power-law is essential only for those of the lowest isolation status
  (highest density regions).
\item The characteristic mass and the slope of the main power-law are
  decreasing functions of halo isolation, as expected from the
  gravitational growth of haloes in increasingly dense regions.
\item An unexpected upturn of $M_{\star}$ occurs for the highest halo
  isolations, implying that the most isolated haloes tend to be
  analogously more massive than in less extreme isolation cases. This
   is present only at higher redshifts and disappears at lower
  redshifts, and it indicates an evolution of the isolation status of the
  most isolated relatively high-mass haloes towards lower
  isolations.
\item The halo abundances of the less isolated haloes (i.e., those in
  the densest regions) show a downturn for $M\lesssim 10^{14}
  M_{\odot}$, in accordance with results related to
  ``knots'',  based on various web-element finder algorithms.
\end{itemize}

We plan to use our isolation criterion to study the inter-halo dynamics 
in different environments, which might provide further insight into the
structure and galaxy formation processes, in an attempt to investigate 
the extent to which the local environment is a dominant determining factor
of physical processes.

\acknowledgements
PSC is supported by the European Research Council under the European
Community's Seventh Framework Programme (FP7/2007-2013 Grant Agreement
no. 279954). We also thank Kostas Karpouzas for his invaluable help in
using the MCMC code.


\end{document}